# Absence of superconductivity in pulsed laser deposited Au/Ag modulated nanostructured thin films

Abhijit Biswas[1], Swati Parmar[1], Anupam Jana[2], Ram Janay Chaudhary[2] and Satishchandra Ogale[1,*]

[1]Department of Physics and Centre for Energy Science, Indian Institute of Science Education and Research (IISER) Pune, Dr. Homi Bhabha Road, Pune, Maharashtra-411008, India
[2]UGC-DAE Consortium for Scientific Research, Indore-452001, India

**Correspondence:** satishogale@gmail.com, satishogale@iiserpune.ac.in

**Abstract:**
We have grown Au/Ag modulated nanostructured thin films by pulsed laser deposition using metallic targets. The growths performed at room temperature involved deposition of alternate layers of the two metals with larger thickness of Au and much lower thickness of Ag (two cases) on crystalline silicon (100) and quartz substrates. However, the characterization revealed that nanostructured configuration is realized in the film. We characterized the films by x-ray diffraction (XRD), Field emission scanning electron microscopy (FESEM) with energy dispersive analysis of x-rays (EDAX), magnetization (M-T) and resistivity ($\rho$-T) measurements. The magnetization and resistivity measurements confirm that there is no signature of superconducting transition within the temperature range of $5K \leq T \leq 300K$ studied in these films.

**Introduction:**
Following the recent report about the observation of possible high temperature and potential room temperature superconductivity by Thapa and Pandey,[1] and subsequent observation regarding similarity of noise patterns in the plot by Skinner,[2] there has been an intense debate in the scientific community, in social media and even some reports in journals[3], about the validity of the claim and other related issues, including the possibility of percolation as being the cause of the reported observations without the need to invoke superconductivity.[4] A recent article by Bhaskaran has also proposed the possibility of superconductivity in monovalent metals and interfaces.[5]

When we read the first report by Thapa and Pandey,[1] we realized that it may be non-trivial to realize the specific state of the sample discussed in their work without clear and detailed information about the specific steps in the synthesis. Since their material is stated to be an Ag incorporated in Au configuration, we thought of making such Au/Ag modulated configuration by another method that can be easily reproduced and tested in any lab. Thus, we made depositions of the Au/Ag modulated films by pulsed laser deposition (PLD) on Si and quartz substrates and tested them for their structural, compositional, magnetic and transport properties to search for a possible signature of superconductivity therein. By modulated we mean alternate depositions of layers of Au and Ag layers of specific thicknesses (Ag thickness much lower than Au) by using separate metallic targets. However, two important things need to be mentioned before we look at the results of the measurements on the samples grown. First, there is a possibility of forming a mixed (and not an ideal composition-modulated superlattice-like) nanostructure by this process since many metals tend to ball-up due to wetting issues on the substrate. Second is the possibility of alloy formation. Although the deposition is performed at room temperature, the pulsed laser generated plasma has enough high energy ionic radicals to transfer energy to the surface and induce mixing of atoms in the deposit. Without high-resolution cross-section TEM data (to which we have no access at this time) we would not know the precise nature of composition modulation in the films. Nonetheless the procedure given here can be easily reproduced in any laboratory having a PLD set up. We felt that our design of experiment could also contribute to the potential discussion on the proposal by Bhaskaran[5] at least within the framework on the specified materials synthesis protocol.

**Thin Film Growth:**

We have grown two different sets of Au/Ag thin films on Si and Quartz substrates by PLD (KrF = 248 nm). High purity commercially available Au and Ag targets were used for the depositions. The films were grown at room temperature and in vacuum (~$10^{-6}$ mBar). The laser repetition rate was kept at 5 Hz, with fluence ~2 J/cm$^2$. The target to substrate distance was ~35 mm. During growth, Au and Ag were deposited sequentially while keeping the number of laser pulses fixed for each target. In one set of samples, for Au and Ag, we gave 200 and 20 pulses individually, whereas 200 and 50 pulses were given for another set of samples. Such a duty cycle was repeated for 100 times, giving the total film thickness of ~80-100 nm. Henceforth, we shall refer these samples as following: Au (200)/Ag (20) on Si as S-1; Au (200)/Ag (20) on quartz as Q-1; Au (200)/Ag (50) on Si as S-2; Au (200)/Ag (50) on quartz as Q-2.

**Results and Discussions:**

The x-ray diffraction (XRD) patterns of these films show the film peaks and the substrate peaks, the film peaks being broader suggesting nanostructured material (**Figure 1a**). The film with less Ag shows better orientation as compared to the film with higher Ag content. Since the lattice constants are almost the same for both the elements, Au (4.078 Å) and Ag (4.085 Å), in the XRD pattern the Bragg peaks corresponding to the unit cell of Au and Ag are indistinguishable, as they almost merge together.

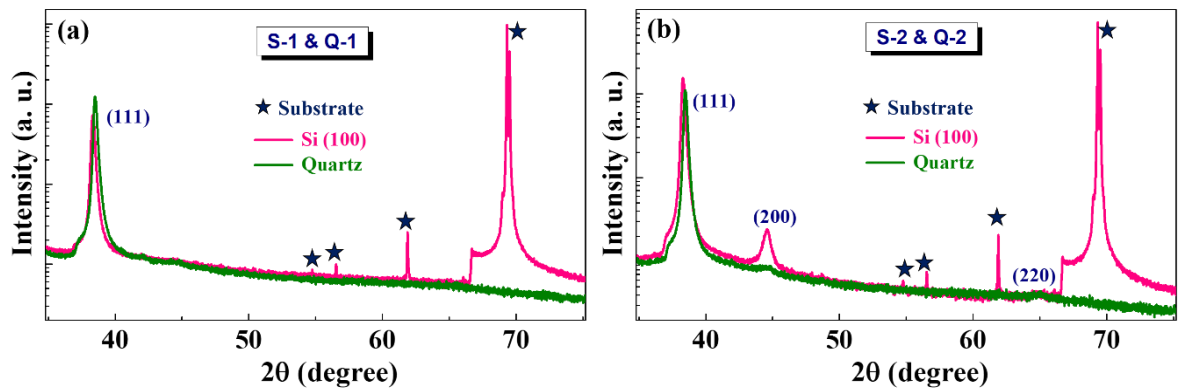

**Figure-1:** XRD patterns of (a) S-1, Q-1 and S-2, Q-2 thin films on Si and quartz substrate. (*) are the substrate peaks.

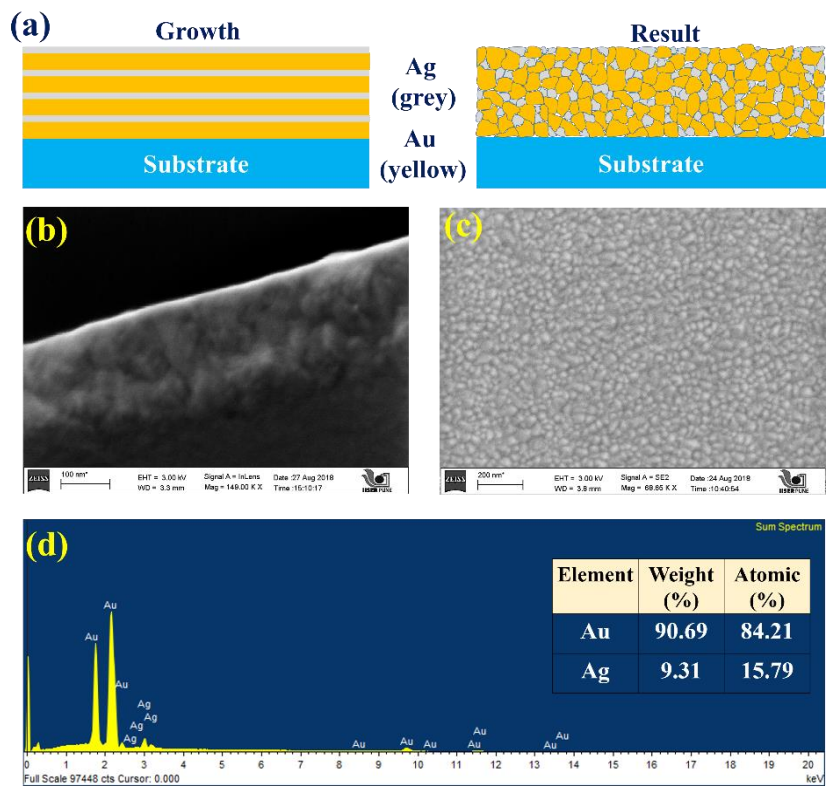

**Figure-2:** Schematics of (a) expected and resultant growth of Au/Ag modulated thin films. (b) Cross-sectional FESEM image and (b) top view and of sample S-1. (d) EDAX shows the elemental mapping and distribution of Au and Ag in the film.

Next, Field Emission Scanning Electron Microscope (FESEM) images were taken for the sample S-1 to check the morphology and composition of the film. It shows the nanostructured nature of the film (**Figure 2**). Compositional analysis by Energy Dispersive X-Ray Analysis (EDAX) shows the preservation of compositional ratio of Au and Ag (80±5:10±5). Moreover, elemental mapping showed that both the elements were present in the film.

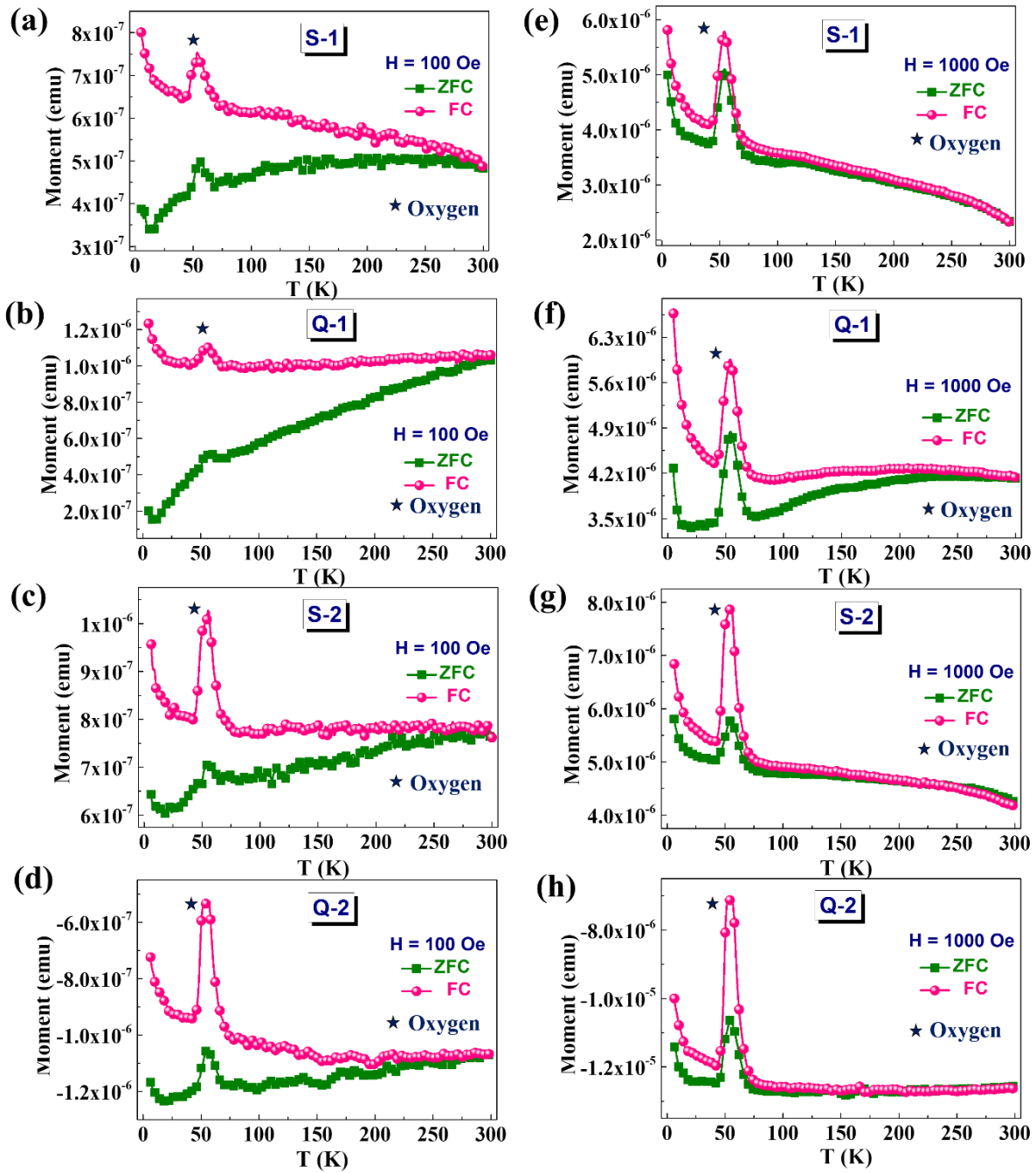

**Figure-3:** Temperature dependent magnetization data for four different samples with the applied magnetic field of 100 Oe (left panel) and 1000 Oe (right panel). Peaks at $T = 50$ K correspond to oxygen contamination from the SQUID.

Now we present the most important results of magnetization measurements. It is known that superconductors expels magnetic flux (the so called Meissner effect, perfect diamagnetism). Thus, at the superconducting transition temperature, if superconductivity exists at all, magnetic measurement must also show a transition (Diamagnetism). To confirm whether our thin films show any signature of superconductivity or not, we searched for the transition in our magnetic measurements which were done by the SQUID magnetometer. Both zero-field cooled (ZFC) and field-cooled (FC) measurements were performed with the applied magnetic fields of 100 Oe and 1000 Oe. The corresponding data presented in Figure 3 are the data as obtained on the film supported on the substrate, with substrate contribution included. We separately measured magnetism of the substrates as well and after obtaining emu/gm from the weights of the sample, we could get an estimate of the magnetism of the films via subtraction of the substrate contribution. These data are presented in the supporting information below the main article. It must be stated however that this subtraction can never be perfect hence these numbers should not be taken to be quantitatively precise representation of the film property.

The ZFC and FC curves in **Figure 3** show a bifurcation, which has been previously reported in the case of Au nanoparticles.[6] For the important issue of superconductivity at hand, however, this is of secondary importance, hence not discussed in any further details. Most importantly, throughout the temperature range of $5K \leq T \leq 300K$, we did not observe any kind of magnetic transition (**Figure 3**), confirming the absence of any plausible superconductivity, at least in our thin films, based on the specified growth procedure.

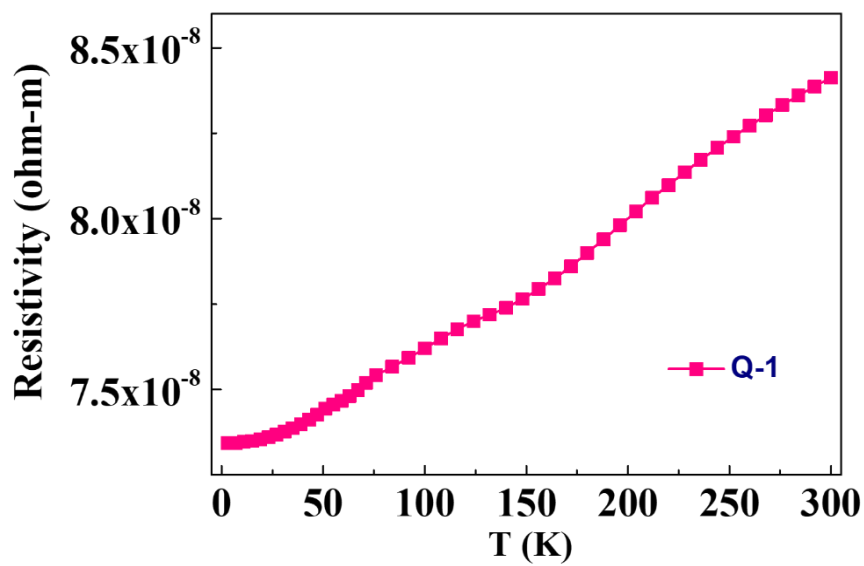

**Figure-4:** Resistivity of sample Q-1. It shows fully metallic behavior throughout the whole temperature range, without any superconducting transition.

Finally, we show the resistivity measurement on a typical sample Q-1 in the set performed by the four probe method in Quantum Design PPMS. Resistivity at $T$ = 300 K was found to be ~$8.4\times10^{-8}$ Ohm-m (almost in the range of the typical values reported for Au and/or Ag). The sample was fully metallic down to the lowest temperature. Resistivity never reached zero, or did not show any kind of transition (downward drop) throughout the measured temperature range of 2K ≤ $T$ ≤ 300K (**Figure 4**), which gives us further confirmation about the absence of superconductivity in our modulated Au/Ag nanostructured thin films.

**Conclusions:**

All the results presented here show that we do not find any signature of superconductivity in pulsed laser deposited Au/Ag modulated nanostructured thin films. We must emphasize that since the sample state is entirely different in this case *vis a vis* what has been reported by Thapa and Pandey, there cannot be any comparison between the two. For instance, in the case of Thapa and Pandey's material, it is not clear from the presented data whether the metallic components are touching each other or they transport through some organic or inorganic insulator via tunneling or another mechanism. In our case it is a fully metallic highly conducting system. We have a limited objective here to show that at least in the Au/Ag modulated nanostructured thin films made by us by the specific and specified PLD process, we do not find any signature of superconductivity. Since there were some remarks in the debates about some theoretical predictions on possible superconductivity in modulated Au/Ag like systems,[5] we hope that our experimental inputs to this debate may contribute some useful insights.

**Acknowledgments**

Abhijit Biswas would like to thank the government of India for providing the SERB-N-PDF fellowship (PDF/2017/000313). SBO will like to thank DST Nanomission Thematic unit for funding support. We would like to thank Imran Khan, Dibyata Rout and Anil Shetti for help in different measurements. We would also like to thank Prof. Surjeet Singh for providing the PPMS facility and expedited measurement schedule for the R-T measurement. The magnetization measurements were performed at UGC-DAE CSR (Indore).

**References:**

[1] D. K. Thapa and A. Pandey, arXiv:1807:08572 (2018).

[2] B. Skinner, arXiv:1808:02929 (2018).

[3] D. Castelvecchi, Nature **560**, 539 (2018).

[4] D. Pekker and J. Levy, arXiv:1808:05871 (2018).

[5] G. Baskaran, arXiv:1808:02005 (2018).

[6] G. L. Nealson et al., Nanoscale **4**, 5244 (2012).

**Supporting Information:**

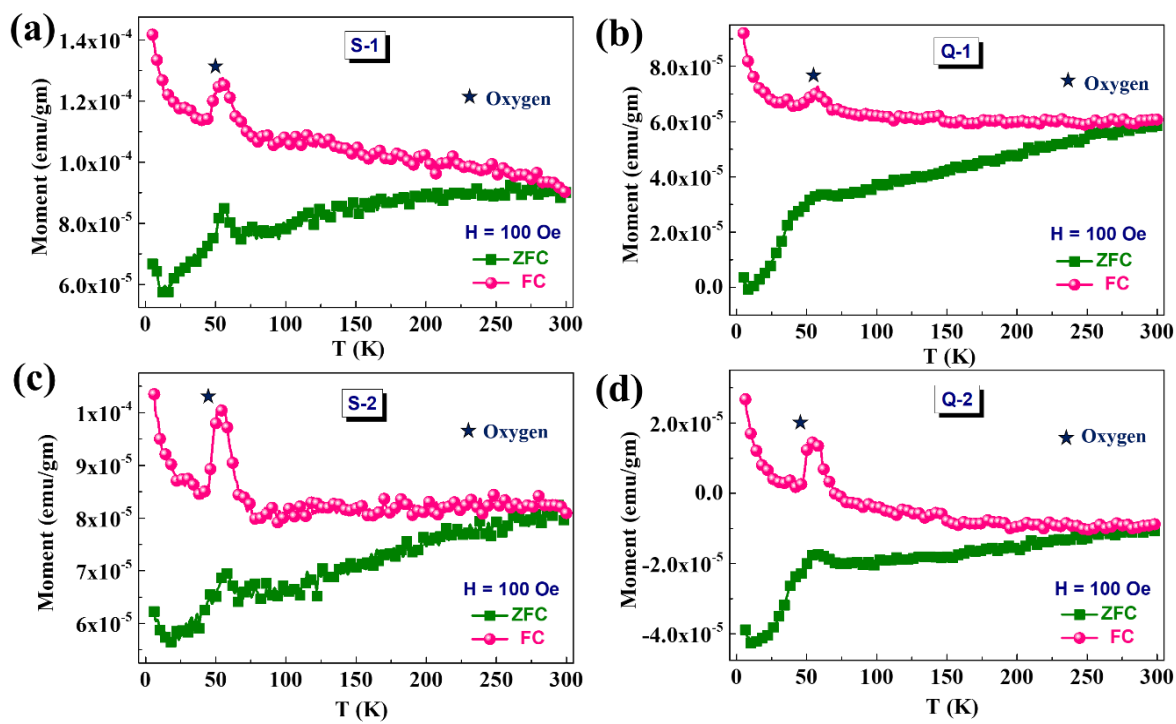

**Figure-S1:** Magnetization of thin films (subtracted from the substrate) with the applied field of 100 Oe.